\documentclass{article}
\usepackage{spconf,amsmath,graphicx,bm}
\usepackage{comment}
\usepackage{cite}
\usepackage{multirow}
\usepackage{color}

\title{MONOCHROME and COLOR POLARIZATION DEMOSAICKING\\USING EDGE-AWARE RESIDUAL INTERPOLATION\footnotemark}

\name{Miki Morimatsu, Yusuke Monno, Masayuki Tanaka, and Masatoshi Okutomi}
\address{Tokyo Institute of Technology, Tokyo, Japan}

\begin{document}
\ninept

\maketitle

\renewcommand{\thefootnote}{}
\footnotetext{\copyright 2020 IEEE. Personal use of this material is permitted. Permission from IEEE must be obtained for all other uses, in any current or future media, including reprinting/republishing this material for advertising or promotional purposes, creating new collective works, for resale or redistribution to servers or lists, or reuse of any copyrighted component of this work in other works. \\ This paper has been accepted by the 27th IEEE International Conference on Image Processing (ICIP 2020).}
\renewcommand{\thefootnote}{\arabic{footnote}}
\setcounter{footnote}{0}

\begin{abstract}

A division-of-focal-plane or microgrid image polarimeter enables us to acquire a set of polarization images in one shot. Since the polarimeter consists of an image sensor equipped with a monochrome or color polarization filter array~(MPFA or CPFA), the demosaicking process to interpolate missing pixel values plays a crucial role in obtaining high-quality polarization images. In this paper, we propose a novel MPFA demosaicking method based on edge-aware residual interpolation~(EARI) and also extend it to CPFA demosaicking. The key of EARI is a new edge detector for generating an effective guide image used to interpolate the missing pixel values. We also present a newly constructed full color-polarization image dataset captured using a 3-CCD camera and a rotating polarizer. Using the dataset, we experimentally demonstrate that our EARI-based method outperforms existing methods in MPFA and CPFA demosaicking.

\end{abstract}

\section{Introduction}
\label{sec:intro}

Polarization is a physical property of an electromagnetic wave such as light consisting of perpendicularly oscillating electric and magnetic fields~\cite{collett2005field}.
Many studies have shown that polarization parameters, such as the angle of polarization~(AoP) and the degree of polarization~(DoP), are valuable for various image processing and computer vision applications, such as specular removal~\cite{jospin2018embedded}, reflection separation~\cite{wieschollek2018separating}, and 3D shape reconstruction~\cite{cui2017polarimetric}, to name a few. 

Polarization images refer to a set of images acquired with different polarizer angles, from which polarization parameters are calculated for every pixel. Polarization images are typically captured by sequentially rotating a linear polarizer placed in front of a camera~\cite{tyo2006review}. However, this conventional approach is not suitable for dynamic scenes and video acquisition, since it requires multiple shots for capturing a set of images. 

\begin{figure*}[t!]
\centering
\centerline{\includegraphics[width=18cm]{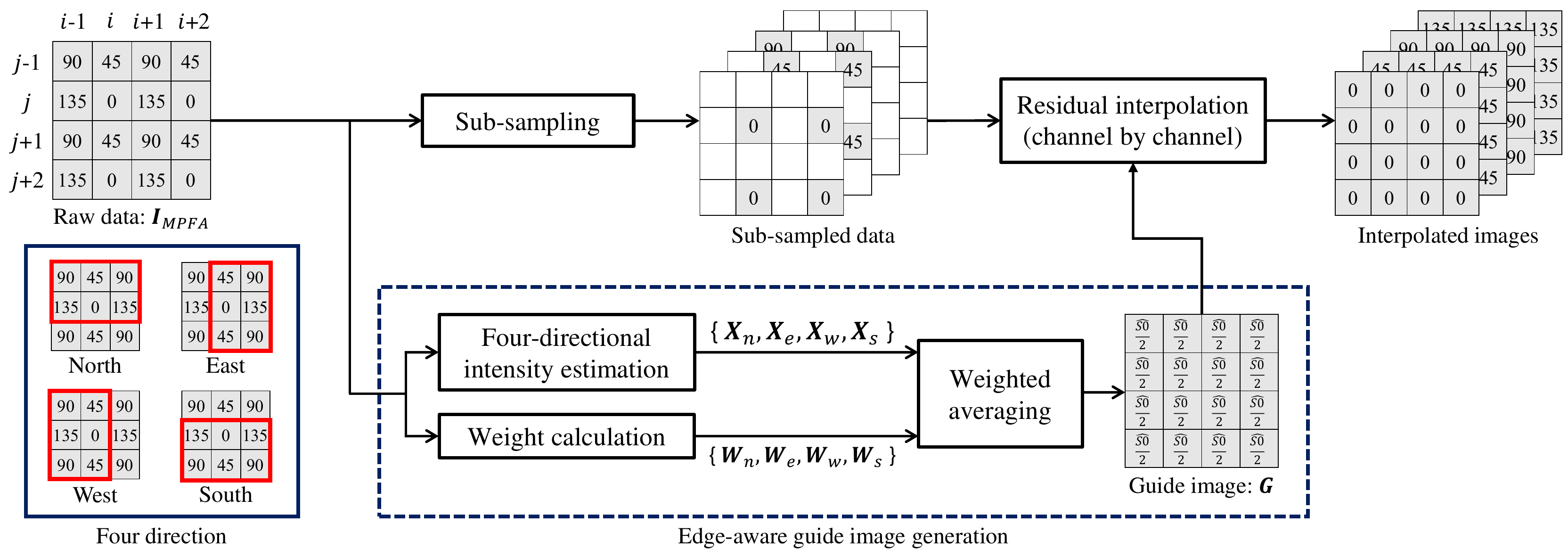}}
\vspace{-3mm}
\caption{The outline of our proposed MPFA demosaicking method based on EARI}
\label{fig:polarizationdemosaicking}
\end{figure*}

As another approach, a division-of-focal-plane or microgrid image polarimeter acquires polarization images by using an image sensor equipped with a polarization filter array~(PFA)~\cite{mihoubi2018survey}. A typical monochrome PFA (MPFA) consists of a 2$\times$2 periodical pattern of four polarizers with angles of 0, 45, 90 and 135 degrees, respectively (see~Fig.~\ref{fig:polarizationdemosaicking}). Similarly, an image sensor equipped with the so-called quad-Bayer color PFA~(CPFA, see Fig.~\ref{fig:outline}) is recently productized by Sony with a much-reduced price from existing color-polarization sensors~\cite{maruyama20183}. These PFA-based sensors are suitable for dynamic scenes and video capturing, since they enable one-shot acquisition of monochrome or color polarization pixel values. For the PFA-based sensors, the demosaicking, which is an interpolation process of missing polarization pixel values, is the key component in acquiring high-quality polarization images.

Many demosaicking methods have been proposed for MPFA, including interpolation-based~\cite{ratliff2009interpolation,gao2011bilinear,gao2013gradient,zhang2016image,li2019demosaicking,ahmed2017residual,ahmed2018four,jiang2019minimized}~(see~\cite{mihoubi2018survey} for a survey), dictionary-based~\cite{zhang2018sparse}, and deep learning-based methods~\cite{zhang2018learning}. Recently, a few methods have also been proposed for CPFA based on the reconstruction-based~\cite{qiu2019polarization} or deep learning-based approach~\cite{wen2019convolutional}. Although learning-based methods~\cite{zhang2018sparse,zhang2018learning,wen2019convolutional} generally achieve higher performance than interpolation-based methods, they are highly data-dependent and require a large amount of training images, which remain difficult to be collected for non-RGB images. They also require high conputational time~\cite{zhang2018sparse} or large memory size~\cite{zhang2018learning,wen2019convolutional}, which is not desirable for integrated sensor systems.

In this paper, we propose a novel MPFA demosaicking method based on residual interpolation~(RI)~\cite{kiku2013residual} and also extend that method to CPFA demosaicking. RI is one of high-performance interpolation methods based on a guide image and has shown its superiority in color demosaicking~\cite{kiku2013residual,kiku2016beyond,ye2015color,kim2015four,monno2015adaptive,monno2017adaptive}. Although RI has also been applied to MPFA demosaicking in some methods~\cite{ahmed2017residual,ahmed2018four,jiang2019minimized,mihoubi2018survey}, they do not fully consider the edge information to generate the guide image. In contrast, we propose a novel \textit{edge-aware} RI~(EWRI) by incorporating a new edge detector to effectively generate a better guide image. 

One limitation of MPFA and CPFA demosaicking research is the lack of evaluation datasets. Although some recent papers have presented the construction of a color-polarization image dataset, it is captured by a Bayer-patterned camera~\cite{lapray2018database,qiu2019polarization} or has not yet published~\cite{qiu2019polarization,wen2019convolutional}. Thus, we constructed a new full 12-channel color-polarization image dataset with 40 scenes\footnote{The dataset and the source code of our proposed method are publicly available at \textcolor{blue}{http://www.ok.sc.e.titech.ac.jp/res/PolarDem/index.html}.} by using a 3-CCD RGB camera and a polarizer rotated with four orientations. Experimental results using the new dataset demonstrate that our EARI-based method outperforms the best-performed existing interpolation-based method in the recent survey paper~\cite{mihoubi2018survey}.

\section{Proposed Method}

\subsection{Proposed MPFA demosaicking method}

Figure~\ref{fig:polarizationdemosaicking} shows the outline of our proposed MPFA demosaicking method based on EARI. We apply RI~\cite{kiku2016beyond}, which is an interpolation method based on a guide image, to interpolate missing pixel values in each polarization orientation. To generate an effective guide image, we propose a novel guide image generation manner considering four-directional intensity and polarization edges. Each processing step is detailed below.

We generate the guide image~${\bm G}$ from raw MPFA data~${\bm I_{MPFA}}$ by estimating four-directional intensity images and averaging them based on the weights for each direction.
By definition, the intensity~$S0$, which is one of the Stokes polarization parameters~\cite{collett2005field}, is expressed as
\begin{equation}
    S0=I_0+I_{90}=I_{45}+I_{135} \,,
\label{eq1}
\end{equation}
where $I_{0}, I_{45}, I_{90}, $ and $I_{135}$ are the associated pixel values of the images at four polarization orientations, respectively.

We explain our idea of four-directional estimation, which is inspired by the horizontal-vertical edge detector of~\cite{zhang2016image}, by taking the intensity estimation at pixel~$(i,j)$ in the north direction (see Fig.~\ref{fig:polarizationdemosaicking}) as example. Based on the definition of Eq.~(\ref{eq1}), for the north region, we estimate the intensity of pixel~$(i,j)$ by two ways as
\begin{equation}
\footnotesize{
\begin{split}
    \widehat{S0}_{n(0,90)}(i,j)=I_0(i,j)+\frac{1}{2} [I_{90}(i-1,j-1)+ I_{90}(i+1,j-1)] \,, \\
    \widehat{S0}_{n(45,135)}(i,j)=I_{45}(i,j-1)+\frac{1}{2} [I_{135}(i-1,j)+ I_{135}(i+1,j)] \,,
\end{split}
}
\end{equation}
where suffix $n$ represents the north region, $\widehat{S0}_{n(0,90)}$ and $\widehat{S0}_{n(45,135)}$ are the estimated intensities using ($I_0,I_{90}$) and ($I_{45},I_{135}$), respectively.
We then calculate the average and the difference of the two estimates as
\begin{eqnarray}
    \widehat{S0}_n(i,j) \hspace*{-0.4em} &=& \hspace*{-0.4em} \frac{1}{2} [\widehat{S0}_{n(0,90)}(i,j)+\widehat{S0}_{n(45,135)}(i,j)],
\label{eq2} \\
    \Delta\widehat{S0}_n(i,j) \hspace*{-0.4em} &=& \hspace*{-0.4em} \widehat{S0}_{n(0,90)}(i,j)-\widehat{S0}_{n(45,135)}(i,j).
\label{eq3}
\end{eqnarray}
If there are no intensity edges and polarization edges~(i.e., edges caused by the polarization parameter differences between pixels) in the region, the difference of Eq.~(\ref{eq3}) becomes zero, meaning that the intensity of Eq.~(\ref{eq2}) is estimated without crossing the edges. Thus, we evaluate the intensity differences for four directions (i.e., north, east, west, and south in Fig.~\ref{fig:polarizationdemosaicking}) to determine the weights of  interpolation directions for generating an edge-aware guide image.

The four-directional intensity estimates and intensity differences can be calculated by filtering the raw MPFA data~${\bm I_{MPFA}}$. The directional intensity estimate~${\bm X}_k = \widehat{\bm{S0}}_k/2$, which is normalized to the pixel value range, is calculated as
\begin{equation}
    {\bm X}_k = {\bm F}_k \otimes {\bm I_{MPFA}}, ~k=\{ n,e,w,s \},
\label{eq4}
\end{equation}
where $\otimes$ represents the filtering operation and $k$ represents the direction of north~($n$), east~($e$), west~($w$), and south~($s$). ${\bm F}_k$ is the filter kernel for $k$-direction, which is expressed as
\begin{equation}
\footnotesize{
\begin{split}
    &\bm F_n = \left[
    \begin{array}{ccc}
      1/8 & 1/4 & 1/8 \\
      1/8 & 1/4 & 1/8 \\
      0 & 0 & 0
    \end{array}
    \right],
    \bm F_e = \left[
    \begin{array}{ccc} 
      0 & 1/8 & 1/8 \\
      0 & 1/4 & 1/4 \\
      0 & 1/8 & 1/8
    \end{array}
    \right],
    \\
    &\bm F_w = \left[
    \begin{array}{ccc} 
      1/8 & 1/8 & 0 \\
      1/4 & 1/4 & 0 \\
      1/8 & 1/8 & 0
    \end{array}
    \right],
    \bm F_s = \left[
    \begin{array}{ccc} 
      0 & 0 & 0 \\
      1/8 & 1/4 & 1/8 \\
      1/8 & 1/4 & 1/8
    \end{array}
    \right] .
\end{split}
}
\end{equation}

The directional intensity difference $\Delta\widehat{\bm{S0}}_k$ is calculated as
\begin{equation}
    \Delta\widehat{\bm{S0}}_k = {\bm H}_k\otimes {\bm I}_{MPFA}, ~k= \{ n,e,w,s \},
    \label{eq:dif}
\end{equation}
where each filter kernel~${\bm H}_k$ is expressed as
\begin{equation}
\footnotesize{
\begin{split}
    &\bm H_n = \left[
    \begin{array}{ccc}
      -1/2 & 1 & -1/2 \\
      1/2 & -1 & 1/2 \\
      0 & 0 & 0
    \end{array}
    \right],
    \bm H_e =  \left[
    \begin{array}{ccc} 
      0 & 1/2 & 1/2 \\
      0 & -1 & 1 \\
      0 & 1/2 & -1/2
    \end{array}
    \right],
    \\
    &\bm H_w = \left[
    \begin{array}{ccc} 
      -1/2 & 1/2 & 0 \\
      1 & -1 & 0 \\
      -1/2 & 1/2 & 0
    \end{array}
    \right],
    \bm H_s = \left[
    \begin{array}{ccc} 
      0 & 0 & 0 \\
      1/2 & -1 & 1/2 \\
      -1/2 & 1 & -1/2 
    \end{array}
    \right].
\end{split}
}
\end{equation}

We then calculate the weight for each direction as
\begin{eqnarray}
    W_k(i,j) &=& \frac{1}{\Delta\widehat{\bm{S0}}'_k(i,j) + \varepsilon}, \label{eq9} \\
    \Delta\widehat{\bm{S0}}'_k &=&
    {\bm M}_k \otimes |\Delta\widehat{\bm{S0}}_k|
\end{eqnarray}
where ${\bm M}_k$ is the 5$\times$5-sized smoothing kernel for $k$-direction, $\varepsilon$ is a small positive value~(set as $10^{-32}$) to avoid the division by zero, and $|\cdot|$ represents the element-wise absolute value operator.

The edge-aware guide image ${\bm G}$ is then generated by the pixel-wise weighted averaging of $X_k(i,j)$ as
\begin{equation}
\footnotesize{
    G(i,j) = \left.　\displaystyle \sum_{k=n,e,w,s} W_k(i,j) X_k(i,j) \middle/ \displaystyle \sum_{k=n,e,w,s} W_k(i,j). \right.
}
\label{eq10}
\end{equation}
Using the generated guide image, we finally apply RI~\cite{kiku2016beyond} to interpolate missing pixel values at each polarization orientation.  

\begin{figure*}[t!]
\centering
\centerline{\includegraphics[width=17.5cm]{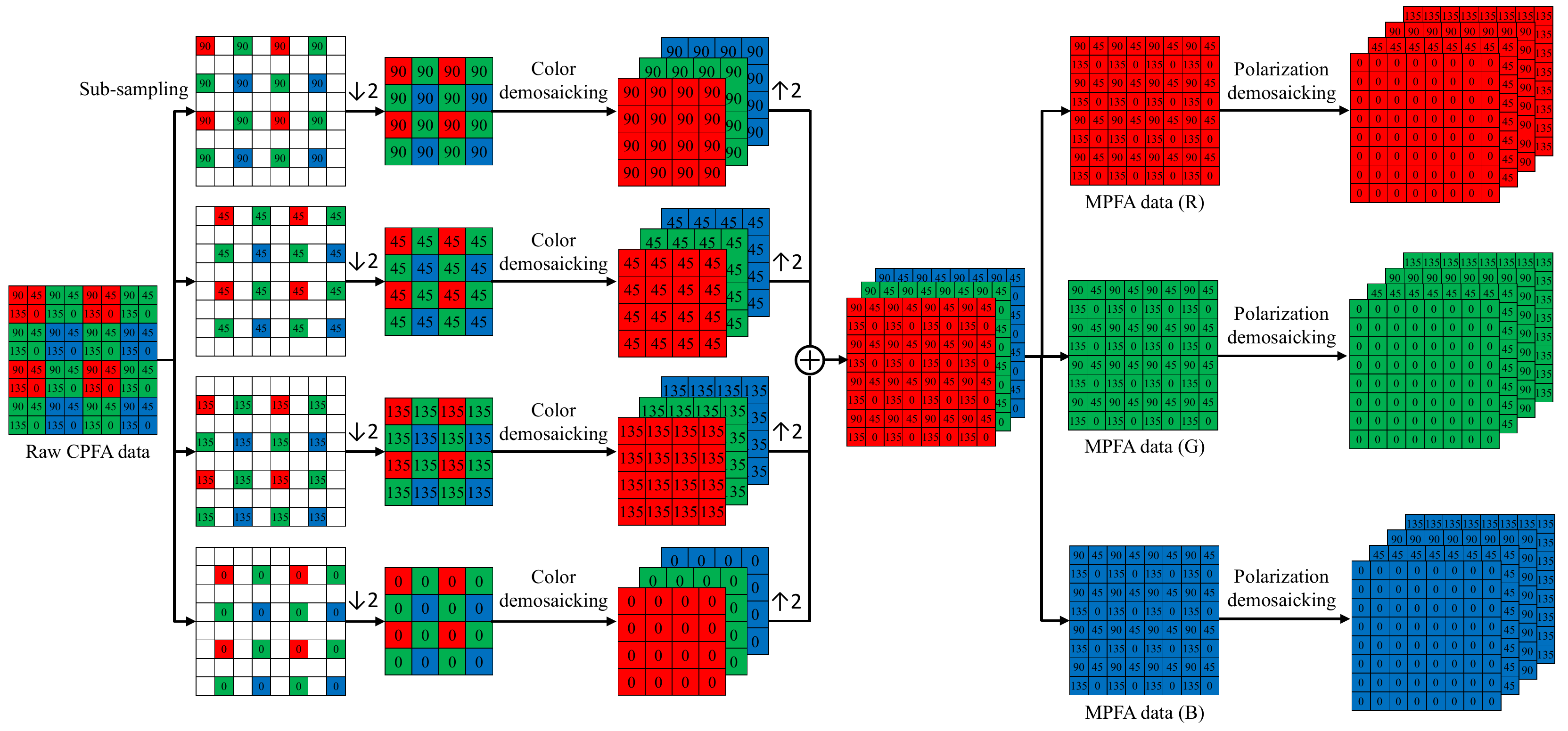}}
\vspace{-4mm}
\caption{The outline of our proposed CPFA demosaicking framework}
\label{fig:outline}
\end{figure*}

\subsection{Extension to CPFA demosaicking}
We here extend our proposed MPFA demosaicking method to CPFA demosaicking. Figure~\ref{fig:outline} shows the outline of our proposed CPFA demosaicking framework, which effectively combines existing color and our MPFA demosaicking methods. We first sub-sample and down-sample~(as expressed by $\downarrow \hspace*{-0.4em}2$) the raw CPFA data to obtain Bayer-patterned data of four orientations. We then apply an existing color demosaicking method to each Bayer-patterned data. The four demosaicked RGB images are then up-sampled (as expressed by $\uparrow \hspace*{-0.4em}2$) and aligned to form the MPFA data of each color channel. Finally, we apply our MPFA demosaicking method to each MPFA data to obtain full 12-channel color-polarization images.

\section{Color-Polarization Image Dataset}
\label{sec:dataset}

We newly constructed the full 12-channel color-polarization image dataset with 40 scenes, as shown in Fig.~\ref{fig:dataset}. Each 12-channel data consists of four RGB images captured with four polarizer angles of 0, 45, 90, and 135 degrees, as shown in Fig.~\ref{fig:dataset}(a). We used JAI CV-M9GE 3-CCD camera, and SIGMAKOKI SPF-50C-32 linear polarizer attached to PH-50-ARS rotating polarizer mount. Each 12-channel data was captured by rotating the linear polarizer placed in front of the camera under an unpolarized light condition. For each polarizer orientation, we captured 1,000 images and averaged them to make the ground-truth image with reduced noise, as performed in~\cite{liu2014practical}. The image resolution is 1024$\times$768 with 10-bit depth.

\section{Experimental results}
\label{sec:result}

\subsection{MPFA demosaicking results}
\label{ssec:monochromeresult}

We evaluated the performance of our MPFA demosaicking method using the green-channel images of our color-polarization dataset as ground-truth monochrome polarization images. We compared our EARI-based method with four interpolation-based methods: bilinear, bicubic, ICPC~\cite{zhang2016image}, and PPID~\cite{mihoubi2018survey}. The PPID method is also based on a guide image called a pseudo-panchromatic image~\cite{mihoubi2017multispectral} and presents the best performance in the recent survey paper~\cite{mihoubi2018survey}.  

\begin{figure}[t!]
\centering
\centerline{\includegraphics[width=7.5cm]{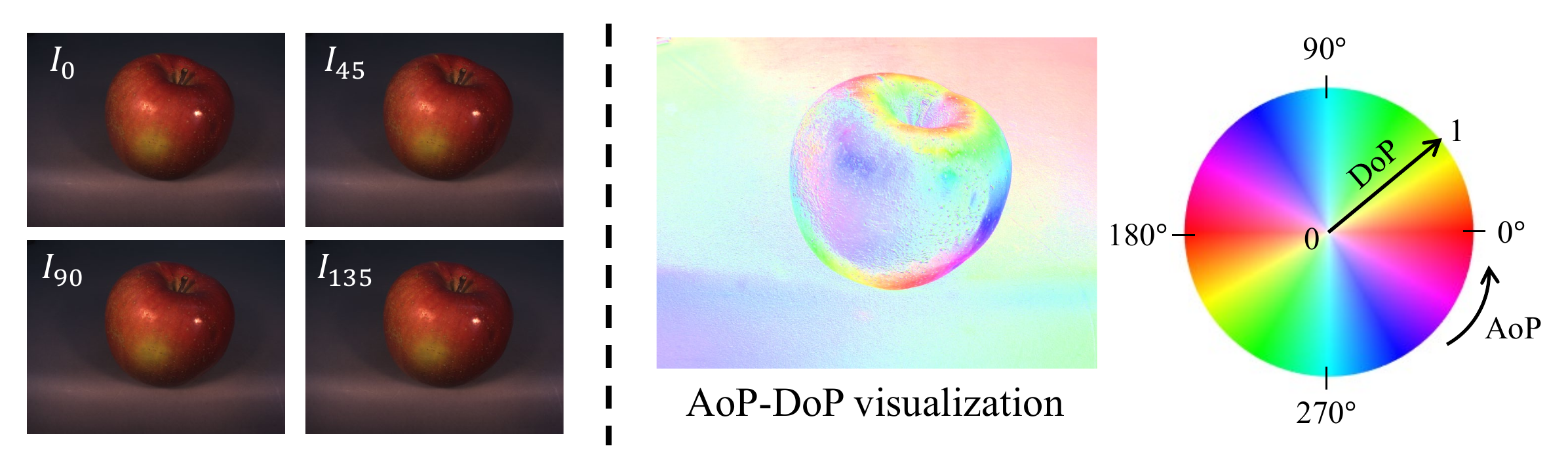}}
\vspace{-1.5mm}
\text{\footnotesize{(a) Example color-polarization image set and its AoP-DoP visualization}} \\ \vspace{2mm}
\centerline{\includegraphics[width=8.5cm]{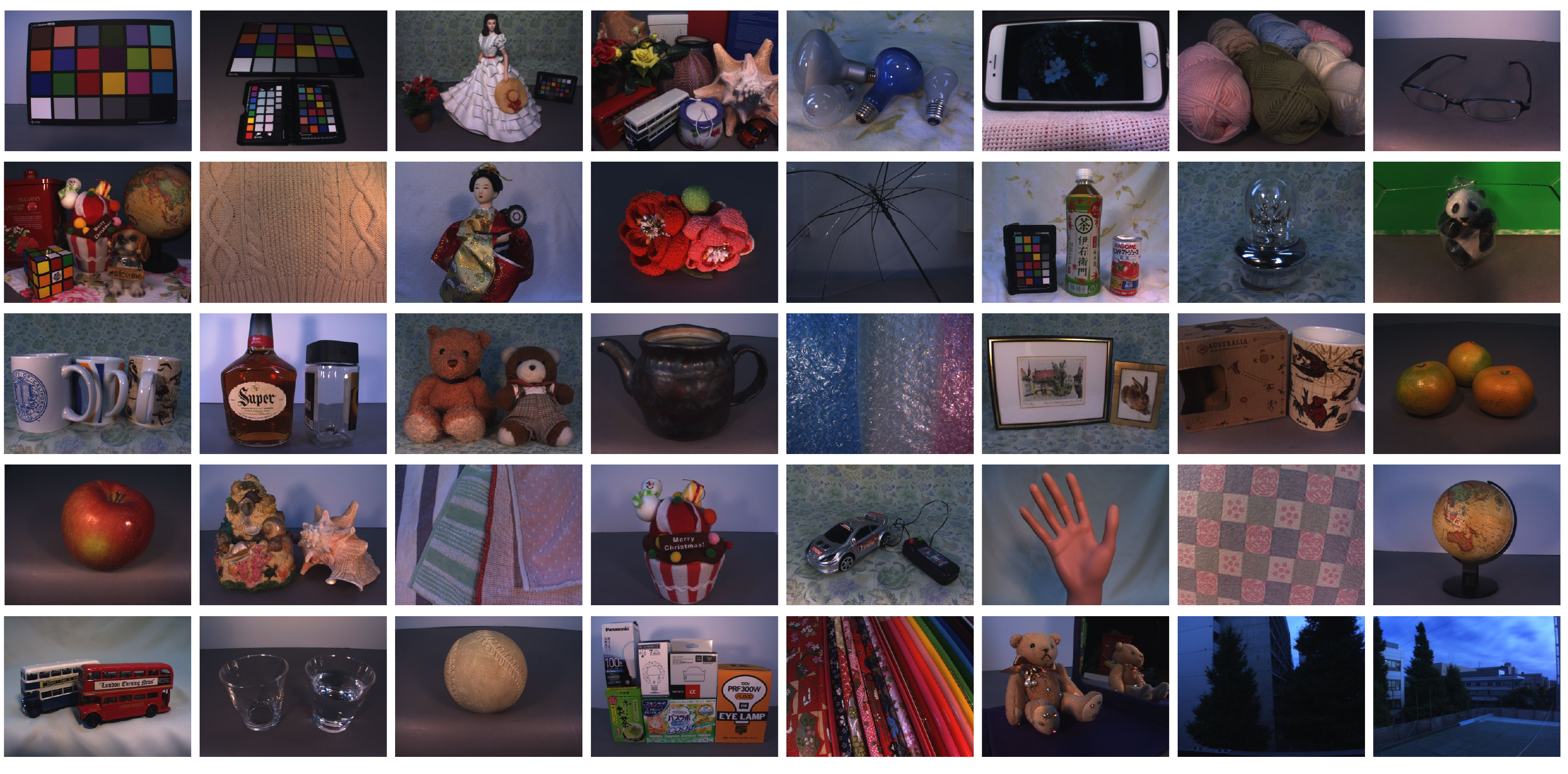}}
\vspace{-1mm}\text{\footnotesize{(b) 40 scenes in the dataset}}\\ \vspace{-3mm}
\caption{Our full 12-channel color-polarization image dataset}
\label{fig:dataset}
\end{figure}

Table~\ref{tab:monochrome} shows the average angle RMSE (RMSE of the angle errors) for AoP images and the average PSNR for four polarization images~($I_{0}$, $I_{45}$, $I_{90}$, and $I_{135}$), three Stokes parameter images ($S0$, $S1$, $S2$), and DoP images. A higher PSNR value and a lower angle error mean better performance. We can see that our EARI outperforms state-of-the-art PPID in most parameters. The bottom row in the table is the RI method using a non-edge-aware guide image generated by a simple non-directional~3$\times$3 averaging filter. We can clearly confirm that, compared with non-directional RI, EARI improves the performance by generating the edge-aware guide image.
Figure~\ref{fig:MPFAresult} shows the visual comparison of selected methods~(see the supplemental material for the results of all methods), where the $S0$ image and the AoP-DoP visualization of Fig.~\ref{fig:dataset}(a) are presented. We can see that our EARI generates a better result without color edge artifact as in ICPC and zipper artifacts as in PPID. 

\begin{figure*}[t!]
\centering
\begin{minipage}{0.28\hsize}
\centering
\includegraphics[width=\hsize]{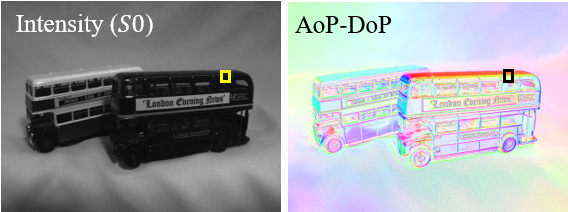}
\end{minipage}
\begin{minipage}{0.083\hsize}
\centering
\includegraphics[width=\hsize]{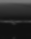}
\end{minipage}
\begin{minipage}{0.083\hsize}
\centering
\includegraphics[width=\hsize]{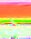}
\end{minipage}
\begin{minipage}{0.083\hsize}
\centering
\includegraphics[width=\hsize]{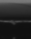}
\end{minipage}
\begin{minipage}{0.083\hsize}
\centering
\includegraphics[width=\hsize]{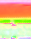}
\end{minipage}
\begin{minipage}{0.083\hsize}
\centering
\includegraphics[width=\hsize]{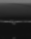}
\end{minipage}
\begin{minipage}{0.083\hsize}
\centering
\includegraphics[width=\hsize]{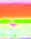}
\end{minipage}
\begin{minipage}{0.083\hsize}
\centering
\includegraphics[width=\hsize]{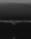}
\end{minipage}
\begin{minipage}{0.083\hsize}
\centering
\includegraphics[width=\hsize]{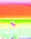}
\end{minipage}
\\
\centering
\begin{minipage}{0.28\hsize}
\centering
\vspace{1mm}
\text{\footnotesize{Full ground-truth images}}
\end{minipage}
\begin{minipage}{0.166\hsize}
\centering
\vspace{1mm}
\text{\footnotesize{ICPC}}
\end{minipage}
\begin{minipage}{0.166\hsize}
\centering
\vspace{1mm}
\text{\footnotesize{PPID}}
\end{minipage}
\begin{minipage}{0.166\hsize}
\centering
\vspace{1mm}
\text{\footnotesize{EARI (Ours)}}
\end{minipage}
\begin{minipage}{0.166\hsize}
\centering
\vspace{1mm}
\text{\footnotesize{Ground truth}}
\end{minipage}

\vspace{-2mm}
\caption{Visual comparison of the intensity image and the AoP-DoP visualization for MPFA demosaicking}
\label{fig:MPFAresult}
\end{figure*}

\begin{table*}[t!]
\centering
\vspace{-3mm}
\caption{Numerical comparison for MPFA demosaicking (Average of 40 scenes)}
\vspace{1mm}
\begin{tabular}{|c||c|c|c|c|c|c|c|c|c|c|}
\hline
\multirow{2}{*}{Method}	&	\multicolumn{8}{c|}{PSNR}	&	Angle error	\\ \cline{2-10}
	&	$I_{0}$	&	$I_{45}$	&	$I_{90}$	&	$I_{135}$	&	$S0$	&	$S1$	&	$S2$	&	DoP	&	AoP	\\
\hline
Bilinear & 42.34 & 41.58 & 42.50 & 41.58 & 44.89 & 46.14 & 45.03 & 33.70 & 21.36 \\
Bicubic	&	43.45	&	42.48	&	43.63	&	42.48	&	46.22	&	47.00	&	45.73	&	34.46	&	20.64	\\
ICPC~\cite{zhang2016image}	&	43.10	&	42.22	&	43.23	&	42.22	&	45.78	&	47.01	&	45.73	&	34.75	&	20.50	\\
PPID~\cite{mihoubi2018survey}	&	46.52	&	44.52	&	46.91	&	44.34	&	48.94	&	50.56	&	47.59	&	\bf{36.96}	&	17.65	\\ \cline{1-1}
EARI (Ours)	&	\bf {47.39}	&	\bf {44.91}	&	\bf {47.84}	&	\bf {44.63}	&	\bf {49.62
}	&	\bf {51.48}	&	\bf {47.83}	&	36.79	&	\bf {17.13}	\\
non-edge-aware 	&	47.01	&	44.68	&	47.41	&	44.46	&	49.27	&	51.02	&	47.70	&	36.25	&	17.44	\\
\hline
\end{tabular}
\vspace{2mm}
\label{tab:monochrome}
\end{table*}

\begin{figure*}[t!]
\centering
\centerline{\includegraphics[width=18cm]{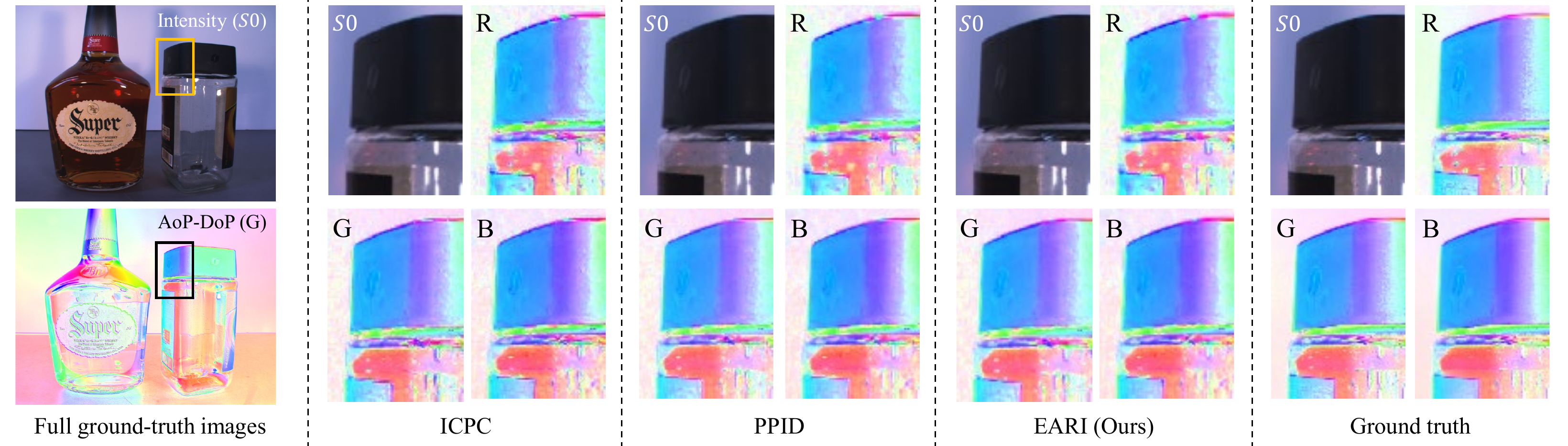}}
\vspace{-3mm}
\caption{Visual comparison of the intensity image and the AoP-DoP visualization for CPFA demosaicking}
\label{fig:colorresult}
\end{figure*}

\begin{table*}[t!]
\centering
\vspace{-3mm}
\caption{Numerical comparison for CPFA demosaicking (Average of 40 scenes)}
\vspace{1mm}
\begin{tabular}{|c|c||c|c|c|c|c|c|c|c|c|c|}
\hline
\multicolumn{2}{|c||}{Method}	&	\multicolumn{8}{c|}{CPSNR}	&	Angle error	\\ \cline{3-11}
\multicolumn{2}{|c||}{(color $|$ polarization)}   &	$I_{0}$	&	$I_{45}$	&	$I_{90}$	&	$I_{135}$	&	$S0$	&	$S1$	&	$S2$	&	DoP	&	AoP	\\
\hline
\multicolumn{2}{|c||}{Bilinear}  &	35.32	&	34.94	&	35.47	&	35.00	&	36.31	&	43.29	&	41.28	&	31.20	&	24.98	\\ \cline{1-2}
\multirow{6}{*}{RI}	&	Bilinear	&	38.34	&	37.79	&	38.50	&	37.86	&	40.03	&	44.09	&	42.58	&	31.96	&	24.15	\\
&	Bicubic	&	38.65	&	38.05	&	38.81	&	38.12	&	40.43	&	44.30	&	42.72	&	32.05	&	24.11	\\
&	ICPC~\cite{zhang2016image}	&	38.61	&	38.01	&	38.77	&	38.09	&	40.33	&	44.49	&	42.87	&	32.36	&	23.86	\\
&	PPID~\cite{mihoubi2018survey} &	39.37	&	38.68	&	39.57	&	38.71	&	40.73	&	46.34	&	44.04	&	\bf{33.98}	&	22.40	\\ \cline{2-2}
&	EARI (Ours)	&	\bf{39.41}	&	\bf{38.72}	&	\bf{39.62}	&	\bf{38.72}	&	\bf{40.76}	&	\bf{46.49}	&	\bf{44.10}	&	33.74	&	\bf{22.18}	\\
&	non-edge-aware	& 39.35	&	38.65	&	39.55	&	38.67	&	40.73	&	46.28	&	43.97	&	33.36	&	22.32	\\
\hline
\end{tabular}
\label{tab:color}
\end{table*}

\subsection{CPFA demosaicking results}
\label{ssec:colorresult}

We next evaluated the performance of CPFA demosaicking using our full 12-channel color-polarization image dataset. We applied minimized-Laplacian RI~\cite{kiku2016beyond} for the color demosaicking step and compared our EARI with the same MPFA demosaicking methods as presented before. We also compared bilinear interpolation applied to each of the sparse 12-channel data as the most basic method.

Table~\ref{tab:color} shows the average color PSNR~(CPSNR) and the average angle RMSE (the average of RGB) for CPFA demosaicking. Similar to the MPFA demosaicking, we can confirm that our EARI provides better performance compared with PPID and also demonstrates the improvement compared with the non-directional RI.
In the visual comparison of Fig.~\ref{fig:colorresult} (see the supplemental material for the results of all methods), our EARI shows the slightly better results on the top of the bin lid, though there still exists jaggy artifacts, since the CPFA demosaickng is very challenging due to the very sparse nature of each color-polarization sample.

\section{Conclusion}
\label{sec:conclusion}
In this paper, we have proposed a novel MPFA demosaicking method based on newly proposed EARI, where we generate the guide image in an edge-aware manner to effectively interpolate missing pixel values without crossing the edges. We also have proposed a CPFA demosaicking framework that effectively combines existing color and our MPFA demosaicking methods. To evaluate the demosaicking performance, we have constructed a new full 12-channel color-polarization image dataset. Using the dataset, we experimentally demonstrate that our EARI-based method achieves better performance than existing methods in both MPFA and CPFA demosaicking. Future work includes the consideration of noise and the proposal of a joint denoising and demosaicking method.

\vspace{3mm}
\noindent
{\bf Acknowledgment.} We would like to thank Dr. Sofiane Mihoubi and Dr. Pierre-Jean Lapray for sharing the excutable code of the PPID method.

\bibliographystyle{IEEEbib}
\bibliography{refs}

\end{document}